\documentclass[12pt,twoside]{article}
\usepackage{fleqn,epsfig,espcrc1}
\usepackage{graphicx}

\def\endauthors{}
\def\authors#1\endauthors{#1}

\def\be{\begin{equation}}
\def\ee{\end{equation}}
\def\br{\begin{eqnarray}}
\def\er{\end{eqnarray}}
\def\brn{\begin{eqnarray*}}
\def\ern{\end{eqnarray*}}
\def\rf#1{{(\ref{#1})}}

\def\lbar{\mbox{$\lambda$\kern-0,450em \vrule width0,35em height1,252ex
depth-1,21ex \kern0,051em}}
\def\dbar{\mbox{d\kern-0,347em \vrule width0,3em height1,252ex depth-1,21ex
\kern0,051em}}
\def\Dbar{\mbox{D\kern-0,735em \vrule width0,3em height0,86ex depth-0,81ex
\kern0,40em}}

\def\qb {{\bf q}}

\def\a {{\alpha}}
\def\b {{\beta}}

\def\g {{\gamma}}

\def\s {{\sigma}}

\def\l {{\lambda}}
\def\m {{\mu}}
\def\n {{\nu}}

\def\N {{{\cal N}}}

\def\ba#1{\begin{array}{#1}}
\def\ea{\end{array}}

\def\be{\begin{equation}}
\def\ee{\end{equation}}
\def\br{\begin{eqnarray}}
\def\er{\end{eqnarray}}
\def\brn{\begin{eqnarray*}}
\def\ern{\end{eqnarray*}}
\def\bit{\begin{itemize}}
\def\eit{\end{itemize}}
\def\bnu{\begin{enumerate}}
\def\enu{\end{enumerate}}
\def\x{\times}
\def\={{\simeq}}

\def\go{\rightarrow  }

\def\rf#1{{(\ref{#1})}}

\def\nn{\nonumber }

\def\2q{{{\{}2{\}}_q}}
\def\3q{{{\{}3{\}}_q}}

\def\qslash{/\!\!\!{q}}
\def\pslash{/\!\!\!{{p}}}
\def\aslash{/\!\!\!{a}}
\def\bslash{/\!\!\!{{b}}}

\begin{document}

\begin{titlepage}
\pagestyle{empty} \baselineskip=21pt
\begin{center}
{\large{\bf One pion production in neutrino-nucleon scattering and the different parametrizations of  the weak $N\rightarrow\Delta$ vertex}}
\end{center}
\authors
\centerline{C. Barbero$^{1,2}$,  G. L\'opez Castro$^{3}$ and A. Mariano$^{1,2}$} \vskip -.05in
\centerline{\small \it
$^{1}$ Departamento de F\'{\i}sica, Universidad Nacional de La Plata, C.
C. 67, 1900 La Plata, Argentina}
\vskip -.05in
\centerline{\small \it
$^{2}$ Instituto de F\'{\i}sica La Plata, CONICET, 1900 La Plata, Argentina}
\vskip -.05in
\centerline{\small \it
$^{3}$ Departamento de F\'{\i}sica, Centro de Investigaci\'on y de Estudios Avanzados,}
\centerline{\small \it
 Apdo. Postal 14-740, 07000 M\'exico, DF, M\'exico}
\endauthors

\bigskip

\centerline{ {\bf Abstract} }
\baselineskip=18pt
\noindent

\bigskip

The $N \to \Delta$ weak vertex provides an important contribution to the one pion production in neutrino-nucleon and neutrino-nucleus scattering for $\pi N$ invariant masses below 1.4 GeV. Beyond its interest as a tool in neutrino detection and their background analyses, one pion production in neutrino-nucleon scattering is useful to test predictions based on the quark model and other internal symmetries of strong interactions.  Here we try to establish a connection between two commonly used parametrizations of the weak $N \to \Delta$ vertex and form factors (FF) and we study their effects on the determination of the axial coupling $C_5^A(0)$, the common normalization of the axial FF, which is predicted to hold 1.2 by using the PCAC hypothesis. Predictions for the $\nu_{\mu} p \to \mu^- p\pi^+$ total cross sections within the two approaches, which include the resonant $\Delta^{++}$ and other background contributions in a coherent way, are compared to experimental data.

\bigskip


{\it Pacs}: 25.30.Pt, 25.75.Dw, 14.20.Gk 13.75.Gx, 14.20.Gk, 13.40.Gp

\vspace{0.5in}

\end{titlepage}

\baselineskip=21pt

\bigskip

Neutrino oscillation experiments search a  distortion in the neutrino flux
at a detector positioned far away (L) from the source. The comparison of near
and far neutrino energy spectra, leads to information about the oscillation probability
$P(\nu_i \rightarrow \nu_j) = sin^2 2\theta_{ij}sin^2{\Delta m_{i,j}^2 L\over 2 E_{\nu}}$,
and then  about the $\theta_{ij}$ mixing angles and $\Delta m_{i,j}^2$ mass squared
differences. Currently, new high quality data are available from
MiniBoone \cite{Mini}, SciBoone \cite{Sci} and new data are expected from Miner$\nu$a \cite{Miner} experiment, which is fully devoted to cross sections measurements of  neutrino-nucleus interactions.

The charged current quasielastic scattering (CCQE) $\nu_l n \rightarrow l^- p$ reaction, with the nucleon bounded in the nucleus target, is usually used as signal event. Although the neutrino energy is not directly measurable, it can be reconstructed from the reaction products through two body
kinematics (exact only for free nucleons). However, competition with other processes
could lead to a possible misidentification of the arriving neutrinos. In fact:

\begin{itemize}
\item{}Disappearance searching experiments $\nu_\mu \rightarrow \nu_x$ (like SciBoone)
use $\nu_\mu n \rightarrow \mu^- p$ CCQE reaction to detect an arriving neutrino and
reconstruct its energy. However, the determination of the neutrino energy  $E_\nu$ could be wrong due to a
fraction of  background events $\nu_\mu p \rightarrow \mu^- p \pi^+$ (CC 1$\pi^+$)
that can mimic a CCQE signal if the pion is absorbed in the target and/or is not detected.

\item{} In $\nu_\mu \rightarrow \nu_e$ appearance experiments (like MiniBooNE) one detects $\nu_e$
in an (almost) pure $\nu_\mu$ beam.
The neutral current reaction $\nu_\mu N \rightarrow \nu_\mu N\pi^0$, $N=n,p$ (NC 1$\pi^0$) can become a source of background
for the signal event $\nu_e n \rightarrow e^- p$  when one of the photons
in the $\pi^0\rightarrow\gamma\gamma$ decay escapes detection leading to a
misidentification of the electron and neutral pion \cite{Ruso10}.

\end{itemize}
Therefore, a precise knowledge of the cross sections of these elementary
\footnote{We refer to neutrino-nucleon scattering as the {\it elementary} process that underlies neutrino-nucleus scattering.} 1$\pi$ processes in  charged (CC) and neutral current (NC) neutrino-nucleon scattering  is a prerequisite
for the proper interpretation
of the experimental data. This will allow to make simulations in event generators to eliminate
fake events coming from 1$\pi$ processes to get more realistic countings of quasielastic (QE) events. We will focus in this work on the CC 1$\pi$ production, which is the channel that enables to fit the axial form factor of our interest.

Several models have been developed over the last thirty years to evaluate the
corresponding elementary cross sections \cite{Fogli79,Hemmert95,Sato03,Hernandez07,Barbero08,Barbero10,Lal82,Leitner09,Lalakulich13,Hernandez13}.
The scattering amplitude in all these models
always contains a resonant term (R)  in the $\pi N$ system, described by the $\Delta(\mbox{1232 })$-pole contribution in Fig.1(h) and (in some cases) by  higher mass   intermediate resonances, plus  a background (B) term describing other processes, as shown in Figs. 1(a)-(f),  (the cross-$\Delta$ contribution in Fig.1(g) can also be included in this background) leading to $\pi N$ final states. Therefore, the scattering amplitude can be written as:  ${\cal M}={\cal M}_{B}+{\cal M}_{R}$.
\vspace{-2cm}
\begin{figure}[h!]
\begin{center}
    \leavevmode
   \epsfxsize =10cm
     \epsfysize = 12cm
    \epsffile{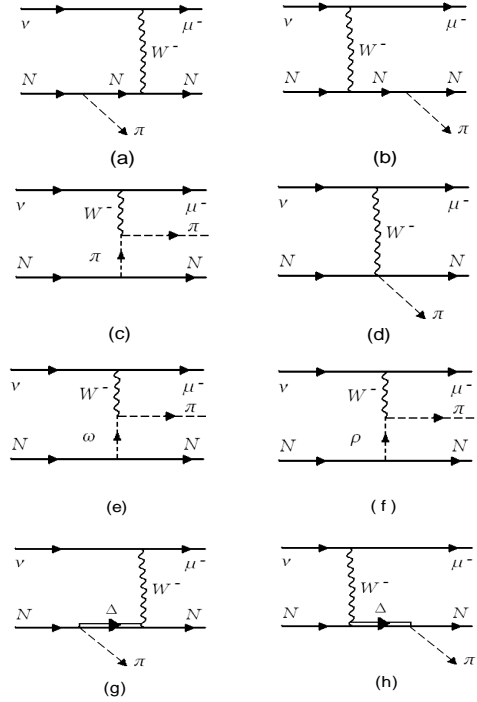}
   \end{center}
\vspace{-2cm}
\caption{Background ((a)-(g)) and resonant (h) contributions to the scattering amplitude.}
\label{fig0}
\end{figure}
Since we are including only the $\Delta(1232)$ as the main resonance contribution, we will compare with data by applying a cut in the $\pi N$ invariant energies at 1.4 GeV.

The difference between all these models stem mainly from the treatment of the  vertexes and the  propagator used to describe the  $\Delta$ resonance and from the consideration (or not) of the background and its interference with the resonant contribution.
 In order to compare the $\Delta$ baryon contribution (both to B and R amplitudes) between different approaches  we need to carefully analyze both, the $\Delta$ propagator and the $\pi N\Delta$ and  $WN \Delta$ vertexes. The
propagator can be written as \cite{elamiri}
\br
G_{\alpha\beta}({\rm p}_\Delta)&=&\frac{\pslash_\Delta+m_\Delta}{{\rm p}_\Delta^2-m_\Delta^2}
\left\{-g_{\a\b}+\frac{1}{3}\g_\a\g_\b+\frac{2}{3m_\Delta^2}{\rm p}_{\Delta\a}{\rm p}_{\Delta\b}-\frac{1}{3m_\Delta}({\rm p}_{\Delta\a} \g_\b - \g_\a{\rm p}_\b)
\right.\nn\\
&-&\frac{b(\pslash-m_\Delta)}{3m_\Delta^2}
\left.\left[\g_\a{\rm p}_{\Delta\b}-(b-1)\g_\b{\rm
p}_{\Delta\a}+\left(\frac{b}{2}\pslash_\Delta+(b-1)m_\Delta\right)\g_\a\g_\b\frac{}{}\right]\right\},
\label{3}\er
with the parameter $b=\frac{A+1}{2A+1}$, where $A$ is an arbitrary parameter related with the contact transformations upon the $\Delta$ field. Since  the physical amplitude should be independent of $A$, the strong and weak vertexes involving the $\Delta$ in Fig. 1(h) should also depend on the $A$-parameter in order  to cancel the $A$-dependence of the corresponding amplitude. In this case both the $\pi N \Delta$ and  $WN \Delta$ vertexes should fulfill these requirements and thus a set of $A$-independent reduced  Feynman rules can be obtained \cite{elamiri}. Equivalently,  one may choose a  common value for A in the Feynman rules involving the $\Delta$ particle to built the amplitude. In Ref. \cite{Barbero08} the value $A=-1/3$ was assumed, coinciding the rules with those in Ref. \cite{elamiri}. However, a common mistake is to use the value $A=-1/3$ which simplifies the vertices simultaneously  with $A=-1$, which simplifies the propagator. This procedure is inconsistent, leading to non-physical expression for the amplitude.

The vector FF's entering the $WN\Delta$ vertex can be fixed from the electromagnetic $\gamma N\Delta$ process by assuming the CVC hypothesis. No analogous symmetry allows to fix the axial-vector FF's.
 Among the axial FF's, the most relevant role is played by $C^A_5(0)$ or, equivalently,  ${D_1(0)=\sqrt{3}}
C^A_5(0)$, depending on the assumed form for the axial vertex at zero momentum transfer. A  reference value is
provided by the PCAC hypothesis being $C_5^A(0)=1.2$ \cite{VonHippel}.  The value $C_5^A(0)\sim 1$ \cite{Hemmert95} is obtained within quark models (QM); however, it is well known that it corresponds to the a `bare' estimate that should be dressed by the pion cloud contribution.
This  dressing can be done dynamically as in \cite{Sato03} where the QM value is enlarged around 35$\%$, or in an effective way by fitting  the experimental data for the
$\nu p\rightarrow \mu^- p\pi^+$ differential cross section \cite{Hemmert95}. Data on weak pion production on  nucleons are scarce and not much precise
 being the most used those obtained by experiments at Argonne National Laboratory (ANL)
\cite{Rad82} and/or Brookhaven National Laboratory (BNL) \cite{Kit86}. The different values assumed or obtained are: $C_5^A(0)=1.20$ \cite{Fogli79}, $1.38$ \cite{Hemmert95,Sato03}, $0.867$ \cite{Hernandez07,Lal82}, $1.35$ \cite{Barbero08,Barbero10}, $1.17$ \cite{Leitner09,Lalakulich13}, $1.00$ \cite{Hernandez13}.
{These different yields depend upon the treatment of the $\Delta$(R+B) contributions to the amplitude as modeled by different authors}. For example in Ref. \cite{Fogli79} the total amplitude is built at tree level by using a complex pole only in the denominator of the $\Delta$-propagator, which is inconsistent with the choice of the  $W$ or $ \pi N \Delta$ vertexes, as it was mentioned above; at the same time, the contributions of Figs. 1(d)-1(g) were not included in the background.
In Ref. \cite{Sato03} the inclusion of pion cloud dynamical effects (PCE) is achieved through a T-matrix approach and all terms are included in the B amplitude, but the same vertex-propagator consistency problems for the $\Delta$ are present.
In Refs. \cite{Hernandez07,Lal82,Hernandez13} the model of Ref. \cite{Fogli79} is extended by adding terms in the B amplitude guided by the effective SU(2) $\sigma$-model Lagrangian, but consistency problems ( $A=-1/3$ in the vertexes and $A=-1$ for the propagator) persist;  a value for $C_5^A(0)$ close to the QM and below the PCAC one is
obtained in this case. In Refs. \cite{Barbero08,Barbero10} the problem of consistency of the $\Delta$ vertex-propagator is solved  together with the question of including the $\Delta$ finite width effects, and  the value obtained for $C_5^A(0)$ is close to the one corresponding to PCE dressed effects. Finally, in Refs. \cite{Leitner09,Lalakulich13}, where  the production and decay of the $\Delta$ resonance are separated in the amplitude, a value close to PCAC is obtained.

Apart from the consistency problems in treating the $\Delta$ resonance, the treatment of the $\Delta$ instability (constant or energy-dependent width) and the adopted convention for the FF's,  the above mentioned models only differ in the way the $WN \Delta$ vertex is parameterized. In view of the different values obtained for $C_5^A(0)$, it would be important to compare these parameterizations.
Let us consider here the amplitude for the elementary neutrino-nucleon CC $1\pi$ production process ($\nu p\go \mu^- p \pi^+$,
$\nu n\go\mu^- n \pi^+$, $\nu n\go \mu^- p \pi^0$). From our Ref. \cite{Barbero08}, hereafter called BLM, we have the total amplitude
\br
{\cal M}_{i}&=&-\frac{G_F V_{ud}}{\sqrt{2}}\bar{u}({\rm p}_\mu)\g_\l(1-
\g_5){u}({\rm p}_\nu)
~~\bar{u}({\rm p}'){\cal O}_i^\lambda({\rm p},{\rm p}',{\rm q})u({\rm p}),
~~i=B,R \label{3p}\er
with $G_F=1.16637\x 10^{-5}$ GeV$^{-2}$, $|V_{ud}|=0.9740$, $({\rm p}$, ${\rm p}_\nu$, ${\rm p}_\mu$, ${\rm k}$, ${\rm p}')$
being the set of 4-momenta of the initial nucleon, neutrino, muon, pion and final nucleon, respectively, and
${\rm q}={\rm p}_\mu -{\rm p}_\nu$ ($Q^2\equiv -{\rm q}^2$) being the momentum transferred from leptons to hadrons.
We adopt here the metric and conventions of Bjorken and Drell (BD) \cite{BD} and for  the hadronic currents $J_{i}^\lambda$ a
vector-axial structure ($J_i^{\lambda}\equiv
V_i^{\lambda} -A_i^{\lambda}$). By assuming the CVC hypothesis in the vector sector, the axial FF at $Q^2=0$ can be fixed from the fit to the  $d\langle \sigma \rangle/dQ^2$ differential cross sections; the strong and other weak couplings involved in ${\cal O}_{B}^\lambda({\rm p},{\rm p}',{\rm q})$ and ${\cal O}_{R}^\lambda({\rm p},{\rm p}',{\rm q})$ are those of the BLM approach. Here we introduce the unstable character of the $\Delta$ by through the complex mass scheme (CMS) \cite{Bar12} consisting  in the replacement $m_{\Delta} \rightarrow m_{\Delta}-i \Gamma_{\Delta}/2$ everywhere the $\Delta$ mass appears in the propagator, with $\Gamma_{\Delta}$ a constant. This procedure avoids the inclusion of  ad-hoc corrections to the vertices in order to restore  gauge invariance  (which occurs if the CMS is adopted only for the denominator of the propagator) in processes where a photon is radiated from the $\Delta$ resonance \cite{elamiri}.

Next, we compare the $W N \Delta$ vertex, defined below as ${\cal W}_{\nu\mu}\equiv{\cal W}^V_{\nu \mu }+{\cal W}^A_{\nu \mu }$, in different prescriptions. Previously, in BLM and \cite{Sato03,Barbero10,Sato96,Mariano07,Scadron} a covariant multipole decomposition analogous to the Sachs choice \cite{Sachs60} of nucleon FF for ${\cal W}^V$ was adopted, namely:
\footnote{We have replaced $q \rightarrow -q$ in Ref. \cite{Sato96} and we have corrected a  misprint (by adding a factor of 2 in the denominator of  $K^M_{\nu \mu}$) in Refs. \cite{Barbero08,Barbero10}.}
\br
{\cal W}^V_{\nu \mu }({\rm p}_\Delta,{\rm q},{\rm p})= \sqrt{2}
\left[(G_M(Q^2)-G_E(Q^2)) K^M_{ \nu \mu}
+ G_E(Q^2) K^E_{\nu \mu}+ G_C(Q^2) K^C_{\nu \mu}\right]. \label{WVDELTA}
\er
The $Q^2$-dependence of FF is assumed to be of the form given  in Ref. \cite{Sato03},  $G_i(Q^2)=G_i(0)(1+Q^2/M_V^2)^{-2}(1 + a Q^2)e^{-bQ^2}\equiv G_i(0)G^V(Q^2)$ with $M_V=0.82$ GeV, $a=0.154/$(GeV$/c)^2$, $b=0.166/$(GeV$/c)^2$.  The Lorentz tensor structures are:
\br
K^M_{\nu \mu}&=& -K^M(Q^2) \epsilon_{\nu \mu \alpha \beta} {({\rm p}+{\rm p}_\Delta)\over2}^{\alpha} {\rm q}^\beta,\nonumber\\
K^E_{\nu \mu}&=&\frac{4}{(m_\Delta - m_N)^2+Q^2}K^M(Q^2)\epsilon_{\nu \lambda \alpha \beta} {({\rm p}+{\rm p}_\Delta)^{\alpha}\over 2} {\rm q}^\beta
\epsilon^\lambda_{\mu \gamma \delta}{\rm p}_\Delta^\gamma {\rm q}^\delta i\gamma_5,\nonumber\\
K^C_{\nu \mu}&=&\frac{2}{(m_\Delta - m_N)^2+Q^2}K^M(Q^2)q_\nu[Q^2 {({\rm p}+{\rm p}_\Delta)_{\mu}\over 2}+{\rm q} \cdot {{\rm p}+{\rm p}_\Delta\over 2}{\rm q}_\mu]i\gamma_5.\label{WGEMC}
\er
with $K^M(Q^2)={3(m_N+ m_\Delta) \over 2 m_N[(m_N+ m_\Delta)^2+Q^2]}$.

Now, we want to express ${\cal W}^V_{\nu\mu}$ in the so-called `normal parity' (NP) decomposition. Using the non-trivial relation \cite{Lein90}
\footnote{The BD convention is used in Ref. \cite{Lein90}.}
\br
- i\epsilon_{\a \b \m \nu} a^\m b^\nu \g_5 = (\aslash \bslash - a\cdot b) i\s_{\a \b}+\bslash
(\g_\a a_\b - \g_\b a_\a)-\aslash(\g_\a b_\b - \g_\b b_\a)+(a_\a b_\b - a_\b b_\a),\nn
\er
and assuming a real $\Delta$ as in Ref. \cite{Scadron}, and thus the validity of the $\Delta$ on-shell constrains  (i.e. $\bar{\psi}_\Delta^\mu \g_\mu\simeq0$, $\bar{\psi}_\Delta^\mu p_{\Delta,\mu}\simeq0$,
$p_\Delta^2\simeq m_\Delta^2$ being $\psi_\Delta^\mu$ de $\Delta$ field) we get a simplified version
\br &&{\cal W}^V_{\nu \mu }({\rm p}_\Delta,{\rm q},{\rm p})= \sqrt{2}i
\left\{-(G_M(Q^2)-G_E(Q^2))m_\Delta K_M(Q^2)H_{3 \nu \mu}\right.\nn\\
&+&\left.[G_M(Q^2)-G_E(Q^2) + 2{2G_E(Q^2)(q\cdot p_\Delta)- G_C(Q^2)Q^2\over (m_\Delta-m_N)^2+Q^2}]K_M(Q^2) H_{4 \nu \mu}\right.\nn\\
&-&\left.[2{2G_E(Q^2)m_\Delta^2+ (p_\Delta\cdot q) G_C(Q^2)\over (m_\Delta-m_N)^2+Q^2}]K_M(Q^2) H_{6 \nu \mu}\right \}\g_5,\label{WVDELTAHNV1} \er
where
\br
H_3^{\nu \mu}(p,p_\Delta,q)&=& g^{\n\m} \qslash - q^\n \g^\m ,\nn\\
H_4^{\nu \mu}(p,p_\Delta,q)&=& g^{\n\m} q.p_\Delta - q^\n p_\Delta^\m ,\nn\\
H_5^{\nu \mu}(p,p_\Delta,q)&=& g^{\n\m} q.p - q^\n p^\m ,\nn\\
H_6^{\nu \mu}(p,p_\Delta,q)&=& g^{\n\m} q^2 - q^\n q^\m .\label{oper}\er
Note that $H_5^{\nu\mu}$ tensor does not contribute to Eq. \rf{WVDELTAHNV1}, but it will appear in forthcoming expressions.
Eqs. \rf{WGEMC} are independent of taking $p=p_\Delta\pm q$ (here the $+$ sign corresponds to the $\Delta$-pole contribution (Fig. 1(h))
and $-$ sign to the cross-$\Delta$ term (Fig. 1(g))) which is clear since $\epsilon_{\nu \mu \alpha \beta}q^{\a}q^{\b}=0$. Thus, Eq. \rf{WVDELTAHNV1} is valid in both cases, but the specific value of $q\cdot p_\Delta\ (=\pm{m_N^2 +Q^2-m_\Delta^2\over 2})$ depends on the particular contribution to the amplitude. Now, if we set on the $\Delta$-pole contribution and replace $p=p_\Delta+q$ we can rewrite \rf{WVDELTAHNV1} as
\br
&&{\cal W}^V_{\nu \mu }({\rm p}_\Delta,{\rm q},{\rm p=p_\Delta+q})=i\Gamma^V_{\nu \mu }({\rm p}_\Delta,{\rm q}),\nn\\
\Gamma^V_{\nu \mu }({\rm p}_\Delta,{\rm q})&=& \sqrt{3}
\left[-{C_3^V(Q^2)\over m_N}H_{3 \nu \mu} - {C_4^V(Q^2)\over m_N^2} H_{4 \nu \mu} - {C_5^V(Q^2)\over m_N^2} H_{5 \nu \mu} +{C_6^V(Q^2)\over m_N^2}
H_{6 \nu \mu}\right]\gamma_5,\nn\\
\label{WVDELTAHNV2}\er
where we have introduced a new set of FF's
\footnote{Since $C_6^V(Q^2) \sim -{4 G_E(Q^2)m_\Delta^2+ G_C(Q^2)(m_N^2 + Q^2 - m_\Delta^2)\over (m_\Delta - m_N)^2+Q^2}$ we adopt
$G_C((m_\Delta - m_N)^2) = 2 m_\Delta/(m_\Delta - m)G_E ((m_\Delta - m_N)^2)$, in order to avoid kinematical singularities
when $Q^2\rightarrow -(m_\Delta - m_N)^2$ \cite{Scadron}. As $(m_\Delta - m_\N)^2 \cong (0.04 GeV/c)^2$ we assume
$G_C(0)\cong {2 m_\Delta \over m_\Delta - m_N} G_E(0).$}:
\br
C_3^V(Q^2) &=& {m_\Delta\over m_N} R_M \left[G_M(0)-G_E(0)\right] F^V(Q^2)\nn\\
C_4^V(Q^2) &=&-R_M \left[G_M(0)- \frac{3m_\Delta}{m_\Delta -m_N} G_E(0)\right]F^V(Q^2),\nn\\
C_5^V(Q^2) &=&0,\nn\\
C_6^V(Q^2) &=&-R_M{2 m_\Delta \over {m_\Delta - m_N}}G_E(0)F^V(Q^2),\label{cG}
\er
being $R_M = \sqrt{\frac{3}{2}}{m_N\over m_N+m_\Delta}$ and
$F^V(Q^2)={\left(1+\frac{Q^2}{(m_N+ m_\Delta)^2}\right)^{-1} }G^V(Q^2)$.

Using $m_\Delta=1.211$ GeV \cite{Mariano01b}, $m_N=0.940$ GeV and the effective values $G_M(0)=2.97$ and $G_E(0)=0.055$ fixed from photoproduction reactions \cite{Mariano07}, we get
\be
C_3^V(0)=2.02,\hspace{1cm}C_4^V(0)=-1.24,\hspace{1cm}C_5^V(0)=0,\hspace{1cm}C_6^V(0)=-0.24.
\label{const2}\ee

In order to make a numerical comparison with other calculations that use the NP parametrization, we consider Refs. \cite{Hernandez07} (hereafter denoted as HNV) and \cite{Lal82}, which both use the same model. Our hadronic weak vertices defined in Eq. (2) are related with those used in \cite{Hernandez07,Lal82} (where the $W$ boson is considered as an incoming particle) as
\br
{\cal O}_{B(a,b,c,d,e,f)}^\lambda({\rm q})&=&\pm i [j^\lambda_{cc+}|_{NP}(-{\rm q})+
j^\lambda_{cc+}|_{CNP}(-{\rm q})+j^\lambda_{cc+}|_{CT}(-{\rm q})\nn\\
&+&j^\lambda_{cc+}|_{PP}(-{\rm q})+j^\lambda_{cc+}|_{PF}(-{\rm q})],\nn\\
{\cal O}_{B(g)}^\lambda({\rm q})&=&\pm i j^\lambda_{cc+}|_{C\Delta P}(-{\rm q}),\nn\\
{\cal O}_{R}^\lambda({\rm p},{\rm p}',{\rm q})&=&\pm i j^\lambda_{cc+}|_{\Delta P}(-{\rm q}),
\label{2}\er
where the $j^\lambda_{cc+}|_{i}$
 are given in Eq. (51) from HNV. Here the $+$ sign corresponds to the $p\pi^+$ and $n\pi^+$ final state reactions and $-$ to the $p\pi^0$ one, since for the latter the isospin
matrix elements accounts a minus sign with respect to ours. Let us remark that the authors in HNV include the $\rho$ meson contribution through a modification in the contact term but don't do the same for the $\omega$ one. Also,
the expressions for the $H_{3,4,5}^{\nu \mu}$ tensors agree with those given in Eq. \rf{oper}, but a different expression, $H_6^{\nu \mu}=m_N^2g^{\nu\mu}$, is used for the remaining FF. In addition, they use  the same Eq. \rf{WVDELTAHNV2} but with
$C_i^V(Q^2)=C_i^V(0)F_i^V(Q^2)$ being ($Q^2$ in units of GeV$^2$)
\br
F_3^V(Q^2)=F_4^V(Q^2)={1 \over (1+Q^2/m_V^2)^2 }{1 \over (1+Q^2/4\times m_V^2)^2 },\nn\\
F_5^V(Q^2)={1 \over (1+Q^2/m_V^2)^2 }{1 \over (1+Q^2/0.776\times  m_V^2)^2 },
\label{HNVFF}\er
with $m_V=0.84$ GeV and
\be
C_3^V(0)=2.13,\hspace{1cm}C_4^V(0)=-1.51,\hspace{1cm}C_5^V(0)=0.48,\hspace{1cm}C_6^V(0)=0.
\label{const22}\ee
In Eq. \rf{const2} we get $C_5^V(0)=0$ as assumed in the $M_1$ dominance model
\footnote{$C_i^V(Q^2)$ are obtained from photo and electroproduction data of $\Delta$ in
terms of the multipole amplitudes $E_{1+},M_{1+}$, and $S_{1+}$.
Recent data determine that $E(S)_{1+}/M_{1+}\sim - 2.5\%$, and this 'dominance' of $M_{1+}$ leads
to $C_5^V(0)=0$ and the relation $C_4^V(0)=-{m_N\over m_\Delta}C_3^V(0)$ \cite{Ruso98}.}
and  $C_6^V(0)\neq 0$ since our
$H_6^{\nu \mu}$ satisfies current conservation condition ($q_\nu H_6^{\nu \mu}=0$) as demanded by the CVC hypothesis. As it can be observed from Eqs. \rf{const2} and \rf{const22}, our values for $C_{3,4}^V(0)$ are consistent with each other.

Now, let us consider the axial-vector contribution ${\cal W}^A_{\nu\mu}$. Within the BLM model, the axial vertex  is taken as in Refs. \cite{Hemmert95,Sato03}, which can be obtained after multiplying  ${\cal W}^V_{\nu\mu}$ by $-\gamma_5$. It reads
\br
{\cal W}^A_{\nu \mu}({\rm p}_\Delta,{\rm q},{\rm p})&=&i \left[D_1(Q^2)g_{\nu \mu }
-\frac{D_2(Q^2)}{m_N^2}({\rm p}+{\rm p}_\Delta)^\alpha(g_{\nu \mu }{\rm q}_\alpha - {\rm q}_\nu g_{\alpha \mu})+
\frac{D_3(Q^2)}{m_N^2} {\rm p}_\nu {\rm q}_\mu\right.
\nn \\
 &-& \left.i\frac{D_4(Q^2)}{m_N^2}\epsilon_{\mu \nu \alpha \beta} ({\rm p}+{\rm p}_\Delta)^{\alpha} {\rm q}^\beta
\gamma_5\right].
\label{WADELTA}\er
The last term in Eq. \rf{WADELTA} will be dropped since we will not
take into account the contribution of the $\Delta$ deformation to the axial current, {\it i.e.}, we set $D_4(Q^2)=0$ and again we use the approximation where the
$\Delta$ is treated as real in the weak vertex, getting
\br
{\cal W}^A_{\nu \mu }({\rm p}_\Delta,{\rm q})\equiv &=&i \left[\left(D_1(Q^2)\pm{D_2(Q^2)Q^2\over m_N^2}\right)g_{\nu \mu }
-\frac{2D_2(Q^2)}{m_N^2}H_{4 \nu \mu}\pm
\frac{D_3(Q^2)+ D_2(Q^2)}{m_N^2} {\rm q}_\nu {\rm q}_\mu\right],\nn\\ \label{WADELTA1} \er
where the $-$ sign corresponds to the weak vertex in Fig. 1(g) and $+$ to that in Fig.1(h).
The $Q^2$ dependence of the FF is \cite{Sato03}
\be
D_i(Q^2)=D_i(0) F^A(Q^2),\hspace{0.1cm}\mbox{for}\hspace{0.1cm}i=1,2,\hspace{0.7cm}
D_3(Q^2)={2 m_N^3\over (m_N+m_\Delta)(Q^2+m_\pi^2)} D_1(0) F^A(Q^2),\label{d00}
\ee
where $F^A(Q^2)=(1+\frac{Q^2}{M_A^2})^{-2}(1 + a Q^2)e^{-bQ^2}$ with $M_A=1.02$ GeV.
The  normalization of the axial FF at $Q^2=0$ is fixed by comparing
the non-relativistic limit of $\bar{u}^\nu_\Delta {\cal W}^A_{\nu \mu} u$ in the $\Delta$ rest frame
($p_\Delta=(m_\Delta,{\bf 0})$, $p=(E_N(\qb), -\qb)$)  with the non-relativistic QM \cite{Hemmert95,Sato03}.
We have
\be
D_1(0) ={6g_A \over\ 5 }{m_N+m_\Delta \over  2 m_N F^A(-(m_\Delta-m_N)^2)},\hspace{1cm}
D_2(0) =-D_1(0){m_N^2 \over  (m_N+m_\Delta)^2},\label{d0}\ee
and we can rewrite
\br
&&{\cal W}^A_{\nu \mu }({\rm p}_\Delta,{\rm q},{\rm p}={\rm p}_\Delta+{\rm q})=i\Gamma^A_{\nu \mu }({\rm p}_\Delta,{\rm q}),\nn\\
{\Gamma }^A_{\nu \mu}({\rm p}_\Delta,{\rm q})&=&\sqrt{3}\left[C_5^A(Q^2)g_{\nu \mu }
-\frac{C_4^A(Q^2)}{m_N^2}H_4^{\nu \mu}+
\frac{C_6^A(Q^2)}{m_N^2} {\rm q}_\nu {\rm q}_\mu\right]. \label{HNVA1} \er
Comparison of Eq. \rf{WADELTA1} (for the plus sign) with  \rf{HNVA1} lead us to the following FF's (note that $C^A_5(0)={D_1(0)\over \sqrt{3}}$)
\br
C_4^A(Q^2)&=&-{2 m_N^2 \over  (m_N+m_\Delta)^2}C_5^A(Q^2)\left[{1 - \frac{Q^2}{(m_N+m_\Delta)^2}}\right]^{-1},\nn\\
C_5^A(Q^2)&=&{D_1(0)\over\sqrt{3}}F^A(Q^2)\left[1 - \frac{Q^2}{(m_N+m_\Delta)^2}\right],\nn\\
C_6^A(Q^2)&=&{2 m_N^3\over (m_N+m_\Delta)(Q^2+m_\pi^2)}C_5^A(Q^2) \left[{(1 - {Q^2+m_\pi^2\over m_N(m_N+m_\Delta)})\over 1 - \frac{Q^2}{(m_N+m_\Delta)^2}}\right].
\label{BLM2}
\er
The corresponding expression from the HNV authors  (by assuming $C_3^A=0$) are
\be
C_4^A(Q^2)=-\frac{1}{4}C_5^A(Q^2),\  \ C_5^A(Q^2)=C_5^A(0) F^A(Q^2),\  \ C_6^A(Q^2)=C_5^A(Q^2){m_N^2 \over Q^2+m_\pi^2},\label{HNV2}
\ee
with $F^A(Q^2)=(1+Q^2/m_A^2)^{-2}(1+Q^2/3\times m_A^2)^{-2}$ and $m_A=1.05$ GeV.
Besides the different dependencies upon $Q^2$ through the $F^A(Q^2)$  functions used in Eqs. \rf{BLM2} and \rf{HNV2}, we observe further differences coming from the
contributions of terms between square brackets in \rf{BLM2}. Note that, at $Q^2=0$, we obtain
\be
C^A_4(0)=-0.38 C^A_5(0),\hspace{1cm}C^A_6(0)=0.87 C^A_5(0){m_N^2\over m_\pi^2},\label{21}
\ee
which are close to the values obtained by HNV, namely
\be
C^A_4(0)=-0.25 C^A_5(0),\hspace{1cm}C^A_6(0)=C^A_5(0){m_N^2\over m_\pi^2}.\label{22}
\ee

Up to now, we have shown that a connection between the Sachs and NP parametrizations of the $WN\Delta$ vertexes
can be established, and that the structure  of the FF {\it under the approximations assumed} are consistent.  Nevertheless, to make complete the comparison,  both models should be confronted within  a numerical calculation where also the fitting of $C^A_5(0)$ enter into the game. We are going to achieve this by using results previously obtained  within the BLM \cite{Barbero08} and HNV \cite{Hernandez07} models.
The effects of adopting different parameterizations for the $Q^2$ dependence of the FF's are shown in Fig. 1, where  we compare the vector FF $F^V(Q^2)$ from BLM with $F^V_{3}(Q^2)=F^V_{4}(Q^2)$ from HNV; the $Q^2$-dependence is shown also for $F^V_{5}(Q^2)$ FF. We also display for comparison, the axial
FF in $C_5^A(Q^2)$ for BLM,  $F^A(Q^2)\left[1 - Q^2/(m_N+m_\Delta)^2 \right]$, and the  corresponding one in HNV model.
As it can be appreciated, we do not expect important differences in the cross sections coming from the different $Q^2$-dependencies of the  FF's. Despite the fact  that  $C_5^V(0)=0$ in the BLM model while  $C_5^V(0)=0.48$ in the HNV one, the magnitude and quick drop of $F^V_{5}(Q^2)$ seems to indicate a small contribution of this form factor.
\begin{figure}[h]
\begin{center}
    \leavevmode
   \epsfxsize =10cm
     \epsfysize = 13cm
    \epsffile{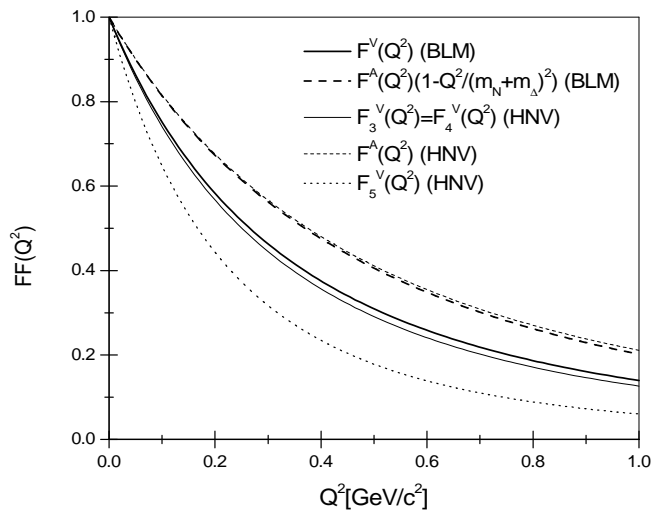}
   \end{center}
\vspace{-6.5cm}
\caption{Comparison of the $Q^2$-dependencies of  the  Vector (V) and axial (A) FF's, between the BLM \cite{Barbero08} and HNV \cite{Hernandez07} models.  The FF  in $C_5^V(Q^2)$ for the HNV model, is also shown for comparison.}
\label{fig1}
\end{figure}
\begin{figure}[h!]
\vspace{-1cm}
\begin{center}
    \leavevmode
   \epsfxsize =11cm
     \epsfysize = 12cm
    \epsffile{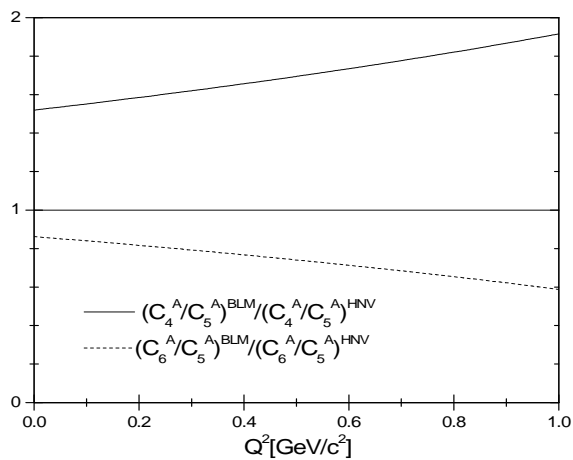}
   \end{center}
\vspace{-6cm}
\caption{The $Q^2$ dependency of the double ratio $C_i^A(Q^2)/C_5^A(Q^2)$ ($i=4,6$) for BLM and HNV models.}
\label{fig3}
\end{figure}

Now, we focus on the effect of the additional $Q^2$-dependent terms  appearing in the $C_{4,6}^A(Q^2)$ FF's in the BLM (Eq. \rf{BLM2}) but not in the HNV (Eq. \rf{HNV2}) model, and also in the normalization conditions, Eqs. \rf{21} and \rf{22} at $Q^2=0$. These effects are better appreciated in the ratio ${C_i^A(Q^2)\over C_5^A(Q^2)}$for  i=4,6 which are displayed in Fig. 3. As it can be observed, the $Q^2$  dependence of these rations is not very strong and the departure from the unity comes essentially from differences in $C^A_{4,6}(0)$. Since the effects of these FF are very suppressed in the cross section  with respect to those due to $C^A_{5}(0)$, we do not expect important differences between both approaches due to these contributions.

Next, we compare calculations for the total cross section of the most relevant $\nu p\rightarrow \mu^- p \pi^+$ reaction, using alternatively the Sachs (Eqs.\rf{WVDELTA},\rf{WGEMC},\rf{WADELTA},\rf{d00} and \rf{d0}) and NP (Eqs. \rf{WVDELTAHNV2},\rf{cG},\rf{HNVA1} and \rf{BLM2}) vertex, within the BLM model. We remark here that, within this model, a value $C_5^A(0)=1.35$ was previously obtained \cite{Barbero08} by fitting the differential cross section $d\langle \sigma \rangle /dQ^2$  using a Sachs decomposition for the weak vertex.

\begin{figure}[h!]
\vspace{2.5cm}
\begin{center}
    \leavevmode
   \epsfxsize =14cm
     \epsfysize = 9cm
    \epsffile{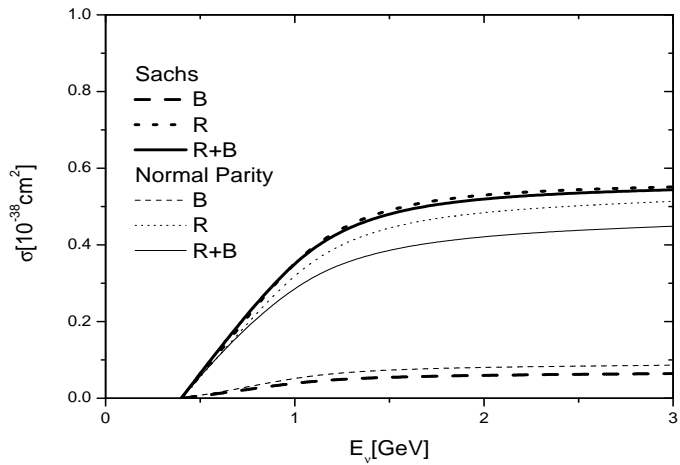}
   \end{center}
\vspace{-6.5cm}
\caption{Total cross section of $\nu p\rightarrow \mu^- p \pi^+$ within the BLM model described in the text. Results using Sachs (thick lines) and Normal Parity (thin lines) decompositions, are shown for the B, R and B$+$R contributions to the scattering amplitude.}
\label{fig2}
\end{figure}

Before we discuss the results, let us mention that the contribution of Fig.1(g) to the $\gamma_0 {\cal W}^{V\dag}_{\mu\nu} \gamma_0$ term (see Eq. \rf{WVDELTA}) appearing in the conjugated amplitude, changes  its sign in the first term of Eq. \rf{WVDELTAHNV1}. Taking into account that $\gamma_0 \left(i H_{3}\g_5,\ iH_{4,6}\g_5\right)^\dag \gamma_0=\left(-iH_{3}\g_5,\ iH_{4,6}\g_5\right) $, the same result is obtained directly from Eq. \rf{WVDELTAHNV2}. Now, the values obtained for $C^V_{4,5}(Q^2)$ are not the same as the ones obtained previously for the 1(h) graph owing to the change of sign for $q\cdot p_\Delta$ in \rf{WVDELTAHNV1} for the cross-$\Delta$ channel. In this sense, the representation given in Eq. \rf{WVDELTA}, apart from the assumed approximations,  is not totally equivalent to that given in Eq. \rf{WVDELTAHNV2}. For the axial part of the cross-$\Delta$ contribution we take into account that
$\gamma_0 \left(ig_{\mu\nu}, iH_{4}, iq_\mu q_\nu\right)^\dag \gamma_0=\left(-ig_{\mu\nu}, -iH_{4}, -iq_\mu q_\nu\right)$ and the minus sign in Eq. \rf{WADELTA1}. We get a different dependence on the $C_5^A(Q^2)$ form factor and the sign of $C_6^A(Q^2)$, but not in the value of $C_5^A(0)$. Again the result will not be the same as taking directly
the  conjugate of Eq.\rf{HNVA1}.

As it can be observed in Fig. 4, results for the resonant R cross section using the NP vertex are slightly below the one obtained by using the  Sachs  vertex for the values of the constants and FF in correspondence. This can be understood considering that moving  from Eq. \rf{WVDELTA} to \rf{WVDELTAHNV2} we have assumed the $\Delta$ to be on-shell (real $\Delta$), which changes the momentum dependence of the vertex, and its coupling to the propagator \rf{3}  that has components behaving differently as  $p_\Delta^2$ increases. As far as the background contribution  B (which includes the graph 1(g)) is concerned, the effect is opposite  and is mainly due to the same approximation, and the effect of the conjugation mentioned above is of minor importance.  As a consequence, the R-B interference will be different in both models  and the cross section obtained within the  NP model will have a value that is below the results obtained using  the Sachs parametrization. This indicates that the fitted value of $C_5^A(0)$ will depend on the specific model used for the weak $WN\Delta$ vertex.
\begin{figure}[h!]
\vspace{-2.cm}
\begin{center}
    \leavevmode
   \epsfxsize =10cm
     \epsfysize = 12cm
    \epsffile{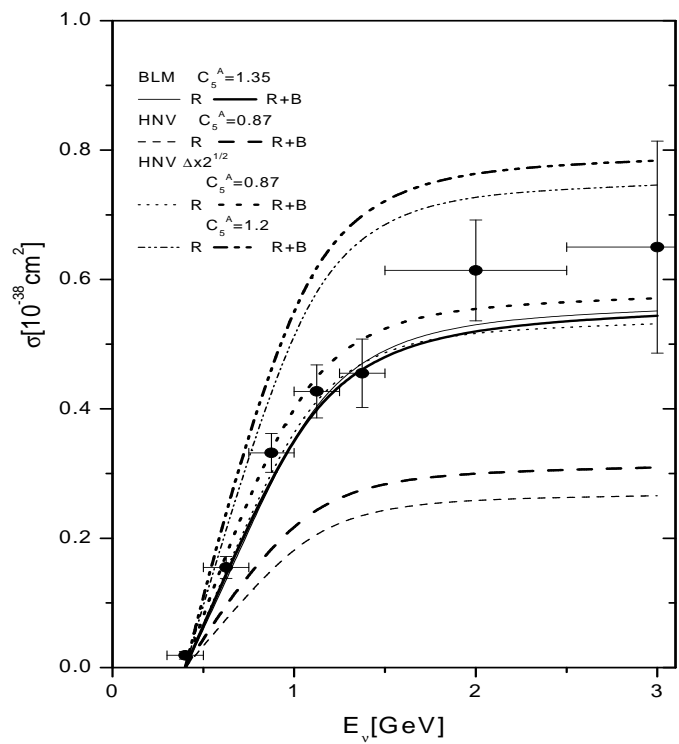}
   \end{center}
\vspace{-2cm}
\caption{The R and R +  B contributions to the cross section in the BLM \cite{Barbero08} and HNV \cite{Hernandez07} models for different values of the axial constant $C_5^A(0)$. Also results for the HNV model  where  $\Delta \times 2^{1/2}$ means that for the $\pi N \Delta$ strong coupling constant we use $f_{\pi N \Delta}/m_\pi \times \sqrt{2} $, are shown both for the same $C_5^A(0)=0.87$ and for $C_5^A(0)=1.2$.}
\label{fig1}
\end{figure}

Finally, we compare the calculations obtained within the BLM model (Sachs form for $WN\Delta$ vertex) with the corresponding ones from HNV (NP form). The main difference between both models, apart from the specific parameters and FF,  is the form adopted for the $\Delta$ propagator in Eq. \rf{3}: we use a value $A=-1/3$ consistent with the adopted for the vertex, and HNV take $A=-1$  which is equivalent to dropping the second term in Eq. \rf{3}. Second, the authors in HNV use an energy-dependent width $\Gamma_\Delta(p_\Delta^2)$, which would need to include energy-dependent vector FF's induced from vertex corrections as it is required by gauge invariance in the case that the corresponding radiative scattering is considered \cite{elamiri,Mariano01b}. We have adopted the value $C_5^A(0)=1.35$ in BLM case \cite{Barbero08} and the value  $C_5^A(0)=0.867$ \cite{Hernandez07} is used for the HNV model (more recently a value $C_5^A(0)=1$ was reported \cite{Hernandez13}).

In Fig. 5 we show results for the $\nu p\rightarrow \mu^- p \pi^+$  total cross section as a function of the neutrino energy $E_{\nu}$;  the
R and R+B contributions are plotted separately. As it can be observed, the results for the resonant R contribution to the cross section in  the  HNV model  (thin dashed lines) roughly account one-half  of the cross section in  the BLM model (thin full lines).  By this reason, we probe with results obtained by using
$C_5^A(0)= 0.867$ and $1.2$ but with  $f_{\pi N \Delta}/m_\pi \times \sqrt{2}$ (which duplicates the R cross section)   within the HNV model, which are shown as ''$\Delta \times 2^{1/2}$''.  The results of these models are compared to experimental  data from Ref. \cite{Kit86} (below an energy cutoff of 1.4 GeV in the $\pi N$ invariant mass).
As it can be observed,  the results of the BLM model agree with data (see also Ref.\cite{Barbero08}); the results from HNV model using  $C_5^A(0)=0.867$ agrees with data only if the resonance  $\Delta$ contribution to the cross section is  multiplied by  a factor of two. Results corresponding to $C_5^A(0)=0.867$ in the HNV model are well below data, and this cannot be attributed to the different parameterizations of the weak vertex (Sachs and NP) since; as we have seen before, these differences are much smaller  if the same value of $C_5^A(0)$ are used. Note also that the results corresponding to $C_5^A(0)=1.2$ and $f_{\pi N \Delta}/m_\pi \times \sqrt{2}$ agree very well with those reported in HNV \cite{Hernandez07} for this value of the axial constant.

In summary, in this work we have compared calculations for the total cross section of the $\nu p\rightarrow \mu^- p \pi^+$ channel by adopting two different prescriptions for the $WN\Delta$ weak vertex. Important differences are observed, showing that the momentum behavior of the Sachs parametrization for the vertex is not the same as the one assumed for the Normal Parity case. As a consequence, the value of $C_5^A(0)$ that is fitted from data depends upon  the specific parametrization of the weak vertex. In our model we use the Sachs parametrization, and make also a comparison with  calculations
adopting the Normal Parity form which get a very different value for $C_5^A(0)$, trying to look for the origin of 
the differences in the weak pion production cross section results.

Acknowledgements:
C.B. and A.M. fellow to CONICET (Argentina) and CCT La Plata, Argentina. GLC is grateful to Conacyt (Mexico) for partial financial support.

\end{document}